\documentclass[conference, a4paper]{IEEEtran}
\IEEEoverridecommandlockouts

\usepackage{layout}
\usepackage{url}
\usepackage{xcolor}
\usepackage{graphicx}
\usepackage[numbers,sort&compress]{natbib} 
\usepackage{booktabs,caption,threeparttable}
\usepackage{algorithm}
\usepackage{multirow}
\usepackage{array,float}
\usepackage{algpseudocode}
\usepackage[left=1.5cm,right=1.5cm,top=1.75cm]{geometry}
\ifCLASSINFOpdf
\else
\fi
\begin{document}
\setlength{\textheight}{695pt}
%
\title{Zone-Based Privacy-Preserving Billing for Local Energy Market Based on Multiparty Computation

}

\author{\IEEEauthorblockN{Eman Alqahtani\IEEEauthorrefmark{1} and
Mustafa A. Mustafa\IEEEauthorrefmark{1}\IEEEauthorrefmark{2}
}
\IEEEauthorblockA{\IEEEauthorrefmark{1}Department of Computer Science,
The University of Manchester, Oxford Road, Manchester, M13 9PL United Kingdom}
\IEEEauthorblockA{\IEEEauthorrefmark{2}imec-COSIC, KU Leuven, Leuven, 3001, Belgium}
Email: eman.alqahtani@postgrad.manchester.ac.uk, mustafa.mustafa@manchester.ac.uk 
\thanks{This work was supported by EPSRC through EnnCore [EP/T026995/1] and by the Flemish Government through FWO-SBO SNIPPET [S007619].}
}

\markboth{Journal of \LaTeX\ Class Files,~Vol.~14, No.~8, August~2015}%
{Shell \MakeLowercase{\textit{et al.}}: Bare Demo of IEEEtran.cls for IEEE Transactions on Magnetics Journals}

\IEEEtitleabstractindextext{%
\begin{abstract}
This paper proposes a zone-based privacy-preserving billing protocol for local energy markets that takes into account energy volume deviations of market participants from their bids. Our protocol incorporates participants' locations on the grid for splitting the deviations cost. The proposed billing model employs multiparty computation so that the accurate calculation of individual bills is performed in a decentralised and privacy-preserving manner. We also present a security analysis as well as performance evaluations for different security settings. The results show superiority of the honest-majority model to the dishonest majority in terms of computational efficiency. They also show that the billing can be executed for 5000 users in less than nine seconds in the online phase for all security settings, demonstrating its feasibility to be deployed in real local energy markets.

\end{abstract}

}

\maketitle

\IEEEdisplaynontitleabstractindextext

\IEEEpeerreviewmaketitle

\section{Introduction}


The use of renewable energy sources (RES) has increased widely, facilitating carbon emissions reduction. Due to the indeterminacy of their output -- which is hard to manage in current markets -- new decentralised energy market models have emerged, known as local energy markets (LEMs). They allow prosumers to trade their excess energy with others in open markets instead of selling it to their contracted suppliers for a feed-in-tariff (FiT) price that is much lower than the retail market prices, thereby enhancing their profits~\cite{Timothy}. 

LEMs typically require their participants to submit bids in advance of the actual trading periods~\cite{Timothy}. Therefore, market participants need to predict the required bid volumes (amount of energy to be traded) based on their historical data and estimated consumption. They are, therefore, prone to errors and hard to be 100\% accurate. Either intentionally or owing to prediction inaccuracy, participants may commit to trade specific volumes of energy but then fail to fulfill their commitments, consequently disturbing the grid stability~\cite{Dudjak}.  

 Different LEM billing models that incentive the market participants to reduce their deviations from their bid commitments have already been proposed~\cite{Madhusudan}. One such model is the billing model with universal cost split where the total deviation cost is split among all market participants (prosumers and consumers). However, this billing model was applied universally for the entire local market area. Different zones of the LEM area may incur larger deviations than others, and the cost of the universal total deviation should be split proportionally. For instance, in one zone, the total deviation might be zero, and the participants within this part should not be accounted for their individual deviations. 

Furthermore, applying this billing model requires utilising individual private information such as individual bid volumes and meter readings. Existing privacy-preserving billing solutions in LEMs propose only payment mechanisms based on bid commitments assuming perfect fulfilment of the committed volumes. Only a limited number of privacy-preserving LEM studies set a mechanism requiring market participants to pay or get paid for the actual amount of
energy they have produced or consumed (measured by smart meter)~\cite{Xiaoyan,Gai,Gaybullaev,Son} or to also account for the energy deviations~\cite{Pop}. However, the trading amount and meter readings of individuals’ real identities are revealed to the network operator or to an independent trusted party.

To address this gap, we propose a novel zone-based privacy-preserving billing protocol considering participants’ deviations based on multiparty computation with different security settings. Specifically, the contributions of this paper are two-fold:
\begin{itemize}
    \item We design a zone-based privacy-preserving protocol for billing allowing suppliers to obtain their contracted customers’ bills while accounting for their customers’ energy volume deviations in LEM and without revealing any of the individual customers’ private data to any party. We use multiparty computation (MPC) to compute the individual bills based on different security settings, namely: passive (semi-honest) and active (malicious) security with an honest majority, and passive and active security with a dishonest majority.

\item We implement and evaluate the computation complexity of our protocol under each security setting to demonstrate its feasibility in real-world settings. 
\end{itemize}

The rest of the paper is organised as follows. Section~\ref{sec:related_work} covers related work. Section~\ref{sec:preliminaries} introduces the preliminaries. Section~\ref{sec:protocol} describes our protocol. Section~\ref{sec:sec-analysis} provides security analysis, while Section~\ref{sec:evaluation} evaluates our protocol. Finally, Section~\ref{sec:conclusion} concludes the paper.

\section{Related Work} \label{sec:related_work}
Security and privacy concerns in local energy markets have been raised in the past~\cite{Mustafa}, and various solutions have already been proposed. A significant number of these solutions are blockchain-based, inheriting its anonymisation feature. Since de-anonymisation is feasible with basic blockchain implementations, assigning fresh pseudonyms for each financial transaction to prevent linkability has been proposed~\cite{Aitzhan,WangY,Dimitriou,Dorri,Baza,Radi,Son}. However, this has been proven insufficient as the link between transactions can be inferred through blockchain analysis~\cite{Reid,Meiklejohn,Shamir,Barber}. To make this analysis less effective and avoid linkability, a decentralised mixing service is deployed in~\cite{Eisele,Aron2}. 

The work proposed in~\cite{Gai} hides sellers' distribution by assigning multiple accounts to each one. For each transaction, financial tokens are allocated dynamically to either one of the sellers' existing accounts or a newly generated account such that they achieve the effect of differential privacy. A similar approach is applied in~\cite{Xiaoyan}, but they aim to protect both sellers and buyers as well as reduce the massive number of accounts generated to hide inactive users.   
 Another line of work utilises verifiable computation schemes such as zero-knowledge proof~\cite{ZhangX2}, blind signature~\cite{Dimitriou}, or both~\cite{Baza,Radi}. Blind signature schemes, for instance, are used to allow a trusted party to create and sign coins for market participants before a trading period so that it does not know the keys behind the coins.

The aforementioned solutions propose privacy-preserving billing models based on the committed volumes by market participants rather than their actual volumes of energy used during the trading periods. Very few studies have considered applying a privacy approach to a billing scheme that assumes imperfect fulfilment of the committed bids or incorporates the deviations in the bills~\cite{Xiaoyan,Gai,Gaybullaev,Son,Pop}. However, the individual trading data are revealed to a trusted party for the payment process. 

In contrast to the previously mentioned solutions, we propose a privacy-preserving billing protocol that is based on the actual consumption/production energy volumes of individuals recorded by their smart meters during the trading periods, takes into account the individual deviations cost in their bills, does not rely on a trusted third party, and protects individual data from all parties. 

\section{Preliminaries} \label{sec:preliminaries}

\subsection{System Model} \label{sec:system_model}
\begin{figure}[t]
\centering
 \includegraphics[width=0.95\columnwidth,trim=4 4 4 4,clip]{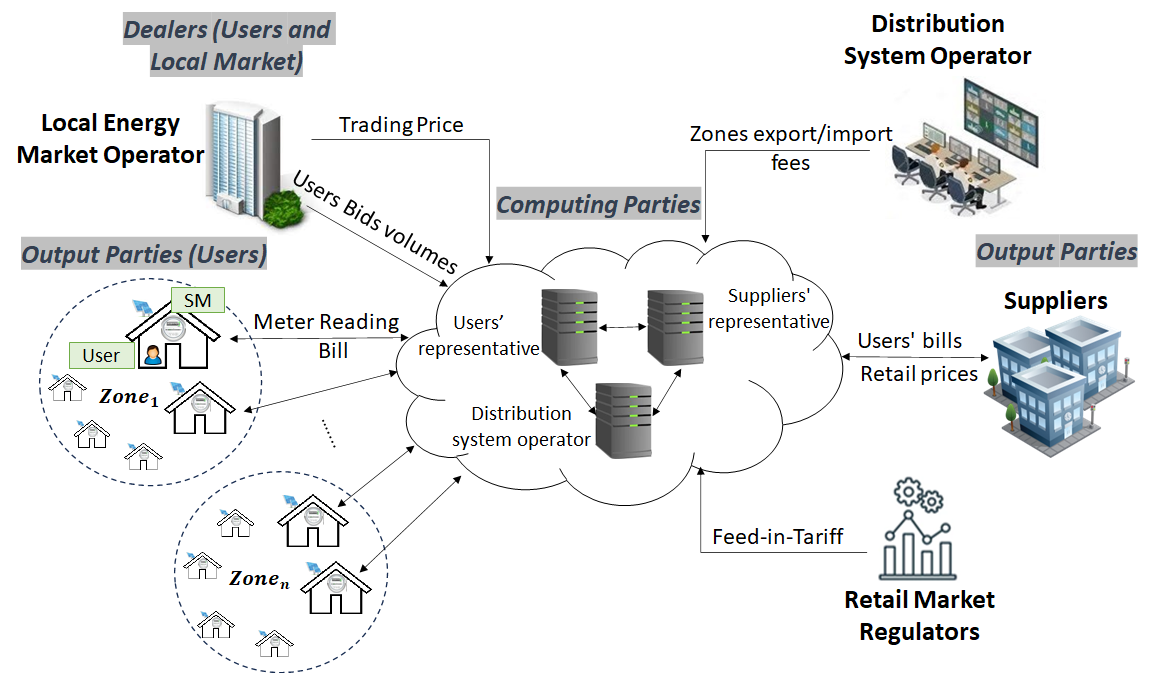}
  \captionsetup{justification=centering}
\caption{System model.}
\label{SystemModel}
\end{figure}

As shown in Fig.~\ref{SystemModel}, our system model consists of the following entities:
\begin{itemize}
\item \textbf{Smart meters (SMs)} are advanced devices that measure the volumes of imported and exported energy by households in nearly real-time and communicate with other entities in the network.
\item \textbf{Users} wish to reduce their bills by participating in a LEM. They submit bids to the LEM to sell their excess energy to others or buy energy at a lower price.
\item \textbf{A Local Energy Market Operator (LEMO)} runs the LEM and determines the trading price and the set of accepted bids to trade for each trading period. 
\item \textbf{Suppliers} provide energy to all users in need. They buy electricity from the wholesale market and sell it to their contracted customers in the retail market at retail prices (determined by suppliers). They are obliged to buy their customers-injected electricity at FiT, which is not traded in the LEM. 
They also issue their monthly customers' bills modified according to their participation in the LEM.
\item \textbf{Retail Market Regulators (RMR)} are entities that set FiT prices users pay to their contracted suppliers for selling energy to them in the retail markets.
\item \textbf{Distribution System Operator (DSO)} manages and maintains the distribution network of a particular area. It divides the LEM area into small zones based on the physical network specifications and historical data that estimate each zone state for each time period. DSO also sets importing and exporting fees for each zone. 
\item \textbf{Computing parties} perform the computations to calculate individual users' bills. 
\end{itemize}

\subsection{Threat Model and Assumptions}\label{Assumptions}
Users, suppliers, LEMO, RMR and DSO are assumed to be malicious. They may try to manipulate users' data for their benefit. Users, for example, may try to modify their own (or other users') bids or meter readings to reduce their bills. Suppliers may attempt to modify users' data to increase their profit. 
All the entities mentioned above may also try to infer individual users' data (i.e., bid volumes and meter readings). Suppliers may want to learn users' bills contracted by other suppliers. External entities are also malicious. They may eavesdrop/modify transmitted data. 

We assume two different settings for the computing parties: \textit{honest majority} (one of the three parties can be corrupted) and \textit{dishonest majority} (two out of the three parties can be corrupted). We further consider two models for each setting: semi-honest and malicious. For the former, the corrupted parties follow the protocol as specified; however, they may try to infer information about users' data (bids, meter readings, and bills). For the latter, the corrupted parties may additionally deviate from the protocol, for example, by sending faulty data during the protocol execution to distort the result. 

Additionally, the protocol is subject to the following assumptions. Every user, supplier, and zone has a unique identifier. SMs are tamper-proof. The communication channels are private and authentic. All entities are time-synchronized.

\subsection{Functional Requirements}
Our protocol should satisfy the following requirements:
\begin{itemize}
    \item Each supplier should learn each of their customers' bills for their participation in the LEM per billing period. 
    \item Each user should learn the individual bill for their participation in the LEM per trading period.
 
\end{itemize}
\subsection{Privacy Requirements}\label{Sec:PRequirements}
Our protocol should satisfy the following requirements:
  \begin{itemize}
      \item Confidentiality: Users' bid volumes and meter readings per trading period should be hidden from all parties. 
      \item Privacy preservation: Exact users' locations, their participation in the LEM, and the type of participation (selling or buying) should be hidden from all parties. 
      \item Authorisation: Users' bills should be accessed only by their contracted suppliers.
  \end{itemize}

\subsection{Multiparty Computation}\label{Sec:MPC}
MPC allows a set of parties to jointly compute a function over private inputs without revealing any data apart from the computation results~\cite{Lindell}. It can be achieved using various cryptographic primitives such as secret sharing, oblivious transfer, and homomorphic encryption. The different primitives can provide either perfect security regardless of adversaries' computation power or computational/conditional security -- a secure protocol given that the adversaries are computationally bounded. Information-theoretic protocols such as BGW can provide perfect security~\cite{Ben-Or}, while protocols that rely on public key primitives such as garbled circuits~\cite{Yao} and SPDZ~\cite{Ivan} can provide conditional security. 

An essential property of MPC protocols is how many parties can be corrupted. 
While information-theoretic protocols provide stronger security, they require an honest majority~\cite{Lindell}. Computational security, on the other hand, can support a dishonest majority; however, they tend to be more complex and expensive~\cite{Hastings}. 
The following are well-known MPC protocols that have been leveraged in our work for each security setting:
\begin{itemize}
\item \textit{Honest-Majority Setting}: We use the optimised secret sharing approach in~\cite{Araki} based on replicated secret sharing specifically designed for three parties and a semi-honest adversary model. The protocol has minimal communication and computation costs as every party sends only one element for each multiplication gate using pseudo-random zero sharing. For the malicious model, we use the protocol proposed by~\cite{Chida} with replicated secrets sharing, which provides security with abort aiming to achieve high efficiency.
\item \textit{Dishonest-Majority Setting}: For the malicious model, we adopt MASCOT protocol~\cite{Keller}. It is an improvement of the original SPDZ protocol, where they replace the expensive somewhat homomorphic encryption used to compute Beaver triples with oblivious transfer. A semi-honest version of the MASCOT protocol can be easily realised by removing all procedures required for malicious security (e.g., MAC generation).
\end{itemize}

\section{Zone-based Privacy-Preserving Billing Protocol for Local Energy Market}\label{sec:protocol}

\subsection{Zone-Based Billing Model with Universal Cost Split} 
 The billing model with universal deviation cost split presented in~\cite{Madhusudan} is modified in order to incorporate users' locations. The LEM area is divided into zones similar to the work presented in~\cite{Baroche}. Then, users' deviation cost are calculated as follows. 
Individual users' deviations per zone are aggregated to calculate each zone's total deviation. Zones' total deviations are then aggregated to calculate the total global deviation for the entire area. If the total global deviation is zero, all users (consumers/prosumers) for all zones pay (get paid) according to the LEM trading price despite their individual deviations. If the total global deviation is positive, then users in the zones with negative and zero total deviation are not accounted for their individual deviations, while prosumers' rewards in the positive total deviation zones are reduced by splitting the cost of the total global deviation among them. 
The cost split is proportional based on the effect each zone had on the global deviation. If the total global deviation is negative, then only consumers in the zones with negative total deviation split the cost of the total global deviation. 

\begin{table}[t]
  \centering
  \footnotesize
  \begin{threeparttable}
  \caption{Notations}
\label{Notations}
\begin{tabular}{p{0.1 \columnwidth} p{0.8 \columnwidth}} 
\toprule
    \textbf{Symbol} & \textbf{Notations} \\
       \midrule
    $tp_k$ & $k$-th time slot, $k \in \{1,2,...,N_k\}$ \\
     $Id_i$ & Unique identifier of user $i, i \in \{1,2,...,N_u\}$ \\ 
    $SId_j$ & Unique identifier of supplier $j$.\\ 
       $ZId_z$ & Unique identifier of zone $z, z \in \{1,2,...,N_z\}$ \\ 
        $N_u^z$ & Number of users who belongs to zone $z$ \\
     $[d]_i$ & Binary value. User $i$ is a seller (1) or buyer (0) during $tp_k$  \\
     $[m]_i$ & Meter reading of user $i$ during $tp_k$  \\
      $[b]_i$ & Bid volume submitted by user $i$ to the LEM for $tp_k$  \\
      $[v]_i^z$ & Individual deviation of user $i$ who belongs to zone $z$ at $tp_k$  \\
      $T$ & Total global deviation \\
       $P$ & Number of prosumers in the entire LEM area \\
        $C$ & Number of consumers in the entire LEM area \\
        $W$ & Zonal deviation weight \\
        $t_z$ & Total deviation of zone $z$ \\
        $p_z$ & Number of prosumers in zone $z$ \\
        $c_z$ & Number of consumers in zone $z$ \\
       $zd_{over}$ & Total deviation of oversupplying zones \\
         $zd_{under}$ & Total deviation of under-supplying zones \\
          $t_i^z$ & Total deviations of zone $z$ to which user $i$ belongs  \\
           $TP$ & LEM trading price during $tp_k$  \\
           $FiT$ & Feed-in-Tariff during $tp_k$ \\
           $RP$ & Retail price during $tp_k$  \\
           $NF_p^z$ & Network fee for exporting in zone $z$ during $tp_k$  \\
          $NF_c^z$ & Network fee for importing in zone $z$ during  $tp_k$ \\

\bottomrule

\end{tabular}
\end{threeparttable}
\end{table}

\subsection{Privacy-preserving Billing Protocol}

Our protocol comprises the following parties:  dealers, computing parties (evaluators), and output parties. Dealers consist of SMs and LEMO. They generate input data shares including bids' volumes (by LEMO) and smart meter readings (by SMs) and send them to the computing parties. Additionally, dealers provide zero input shares for all inactive users to hide who actually participated in the LEM. The computing parties evaluate the MPC function to compute individual users' bills and send the results to users and suppliers. They are three servers with conflicting interests to avoid colluding. We assume that one server is controlled by the suppliers, one by the users, and one by the DSO. The number of servers is chosen to leverage some highly efficient MPC protocols dedicated for three computing parties~\cite{Araki}. We also reduce the high computation and communication cost incurred from having a high number of evaluators when utilising the MASCOT protocol~\cite{Keller}. Output parties are the users and suppliers who receive the resultant bills as shares from the computing parties and reconstruct the outputs. The notation used throughout the paper is given in Table~\ref{Notations}. The square brackets [$x$] denote that $x$ is secretly shared.

Our proposed protocol consists of the following five phases.

    \subsubsection{Generation and Distribution of Input Data}
    Each SM in every zone $z$ and for every trading period $tp_k$ creates a tuple $(Id_i,SId_j,ZId_z,[m]_i)$ which contains shares of its recorded meter reading. Additionally, LEMO generates a tuple for each user $(Id_i,[d]_i,[b]_i)$ consisting of bid volume shares and the state of the user (seller or buyer). SMs and LEMO then send the shares to the computing parties. The applied MPC protocols determine how input data are split into shares. Replicated and additive secret shares are generated for~\cite{Araki,Chida}, and~\cite{Keller}. 

\subsubsection{Zone-based Deviations Aggregation}
       Once the computing parties receive the shares from SMs and LEMO, they first combine the received tuples for each user into one user tuple $ (Id_i,SId_j,ZId_z,[d]_i,[m]_i,[b]_i$). They then proceed to evaluate the total deviation per zone in a data-oblivious fashion as shown in Alg.~\ref{alg:DeviationsAggr}. The parties loop through users' tuples to compute individual users' deviations, total deviation, and number of prosumers and consumers for each zone. The algorithm consists of only additions operations that each party can evaluate locally. The algorithm is executed $N_z$ times, and its output is produced in shared form.
    \begin{algorithm}
      \footnotesize
        \caption{Zone-based Deviations Aggregation}\label{alg:DeviationsAggr}
        \hspace*{\algorithmicindent} \textbf{Input:} Set of $N_u^z$ user tuples $U = (Id,ZId,[d],[m],[b]$)\\
        \hspace*{\algorithmicindent} \textbf{Output:} Zone $z$ tuple $ZN= ([t],[p],[c]$), zone $z$  deviations tuple $D = ([v_0],[v_1],...,[v_{N_u^z}])$ 
        \begin{algorithmic}
        \For{$i = 0$ to $N_u^z$ }
            \State $[v]_i^z \gets [m]_i - [b]_i$
            \State $[t]_z \gets [t]_z + [v]_i^z$
            \State $[p]_z \gets [p]_z + [d]_i$
            \State $[c]_z \gets [p]_z + 1-[d]_i$
        \EndFor
        \end{algorithmic}
        \end{algorithm}

    \subsubsection{Zonal Deviation Weight Computation}\label{SecondPhase}
     The zonal deviation weight $W$ is calculated to help distribute the total global deviation between the zones proportionally. This computation can be done in clear as the required data do not reveal individual users data. This would reduce the overhead of performing comparison and division/multiplication operations. Accordingly, the computing parties first jointly reconstruct the shares of each zone $z$ tuple $ZN= (t_z,p_z,c_z$). Each party then computes the total global deviation by simply summing the total deviation per zone $\sum_{l=0}^{N_z} t_{zl}$. Finally, each party computes the zonal deviation weight locally, as shown in Alg.~\ref{alg:Deviation_Weight}.

            \begin{algorithm}[t]
              \footnotesize
        \caption{Zonal Deviation Weight Computation}\label{alg:Deviation_Weight}
        \hspace*{\algorithmicindent} \textbf{Input:} $T,P,C,$ set of all $t_z$ for $z \in \{1,2,...,N_z\}$ \\
        \hspace*{\algorithmicindent} \textbf{Output:} $W$
        \begin{algorithmic}
        \State $zd_{over} \gets 0$
        \State $zd_{under} \gets 0$
        \If{$T > 0$}
            \For{$z = 0$ to $N_z$ }
                \If{$t_z >0$}
                \State $zd_{over} \gets zd_{over} + t_z $
                \EndIf
            \EndFor
            \State $W \gets \frac{T}{zd_{over}}$  
        \ElsIf{$T <0$}
             \For{$z = 0$ to $N_z$ }
                \If{$t_z <0$}
                \State $zd_{under} \gets zd_{under} + t_z $
                \EndIf
            \EndFor
            \State $W \gets \frac{T}{zd_{under}}$ 
        \EndIf
        \end{algorithmic}
        \end{algorithm}

 \subsubsection{Individual Billing}
     Once the deviation weight is calculated, the computation parties jointly compute individual users' bills for every trading period $tp_k$ (see Alg.~\ref{alg:IndividualBilling}). The parties take as inputs users tuples shares (phase 2), deviation tuples shares computed per zone (phase 2), total global deviation and zonal deviation weight computed (phase 3), and billing prices. The parties loop through the users' tuples to calculate the basic bills using oblivious multiplication and addition operations over the secretly shared meter readings $[m]_i$ and states $[d]_i$. The basic bills are then modified to include the deviation cost after performing oblivious comparisons on individual deviations' shares. 
            \begin{algorithm}
              \footnotesize
        \caption{Individual Billing}\label{alg:IndividualBilling}
        \hspace*{\algorithmicindent} \textbf{Input:} Set of $N_u$ user tuples $U = (Id,SId,ZId_z,[d],[m],[b]$), set of $N_z$ zone deviations tuples $D = ([v_0],[v_1],...,[v_{N_u^z}])$, set of $N_z$ zone tuples $ZN= (t,p,c$), $T,W, TP, NF_p, NF_c, FiT, RP$ \\
        \hspace*{\algorithmicindent} \textbf{Output:} Set of $N_u$ user bills $[bl]_i, i \in \{1,2,...,N_u\}$
        \begin{algorithmic}
        \For{$i = 0$ to $N_u$ }
              \State $[bl]_i \gets [m]_i \times (TP + (-NF_p^z \times  [d]_i) +  (NF_c^z \times (1-[d]_i))) $
            \If{$T >0$}
                \If{$t_i^z >0$}
                \State $[c] \gets [v]_i^z>0$
                \State $[bl]_i \gets [bl]_i + t_z \times \frac{W}{p_z} \times (FiT -TP) \times [c] \times [d]_i$
                \EndIf
            \ElsIf{$T <0$}
                \If{$t_i^z <0$}
                \State $[c] \gets [v]_i^z<0$
                \State $[bl]_i \gets [bl]_i + t_z \times \frac{W}{c_z} \times (RP -TP) \times [c] \times (1-[d]_i)$
                \EndIf
            \EndIf
        \EndFor

\end{algorithmic}
\end{algorithm}

\subsubsection{Distribution of Results} 
For each trading period $tp_k$, the computing parties send the individual bills shares $[bl]_i$ to the corresponding users according to $Id_i$. After a number of trading periods $N_{k}$, the parties aggregate individual bills shares for each user  $\sum_{k=0}^{N_k} [bl]_i^{k}$ and forward the results to their corresponding suppliers according to $SId_j$.

\section{Security Analysis} \label{sec:sec-analysis}
Our assumptions in Section~\ref{Assumptions} imply the security of our protocol against users and external adversaries. In more detail, SMs are assumed to be tamper-proof, which indicates that inputs sent by SMs can not be altered by users. Additionally, we have assumed authentic and private channels which protect against malicious LEMO, DSO, RMR and external adversaries. This can be simply realised using TLS protocol. 

Furthermore, MPC approaches used for our protocol (specifically in Alg.~\ref{alg:DeviationsAggr}~and~\ref{alg:IndividualBilling}) form an arithmetic or a mixed circuit that can be evaluated with no leakage, guaranteeing privacy. Our assembled circuit to execute individual bills function would be as secure as the underlying MPC protocols used~\cite{Canetti}. Therefore, based on MPC, the computing parties have access to only users' input shares and can learn nothing other than what can be inferred from the protocol output. The protocol can be computed with perfect security when utilising replicated secret sharing offering security with an honest majority (one corrupted party) and passive security as in~\cite{Araki} or active adversary as in~\cite{Chida}. Our protocol can also be implemented with a dishonest majority (two corrupted parties) and active or passive adversary by utilising cryptographic primitives such as oblivious transfers in~\cite{Keller}, hence, achieving computational security. As a result, the security of our protocol is derived from the underlying MPC protocols~\cite{Canetti}. 

In addition, suppliers do not receive any of their individual customers' bills for a single trading period -- since inferring individuals' data such as meter readings would be straightforward. Instead, they receive an aggregate of the individual bills corresponding to a number of trading periods. Consequently, we can conclude that the protocols are secure against malicious suppliers and semi-honest or malicious computing parties (based on the underlying MPC protocols).

\section{Experimental Evaluation}\label{sec:evaluation}
\subsection{Implementation Details}
We run the three computational parties on the same machine, a 64-bit Linux server with 16 cores single thread Intel Xeon processors and memory of 64 GB. We executed our experimentation using MP-SPDZ framework~\cite{Keller2}, which supports the underlying primitives and MPC protocols utilised by our protocol (Section~\ref{Sec:MPC}). 
First, we adopt the arithmetic circuits model under which any function consisting of the basic math operations (addition and multiplication) can be constructed and evaluated~\cite{Lindell}. Later, for some security models, we utilise~\cite{Rotaru} to convert from arithmetic to binary computation when evaluating non-linear functions such as comparisons forming what is known as a ``mixed" circuit. 

We adopted the same random data generation mechanism applied in~\cite{Madhusudan} -- based on a realistic dataset used in~\cite{Abidin} -- to simulate bid volumes and meter readings during a trading period. The numbers are represented in Watts so that only integer numbers are assumed. We conducted our experiments starting with 1000 users participating in a LEM for one trading period and gradually increased the number to 5000 users.   

\subsection{Experimental Results}
The underlying MPC protocols used in our protocol divide the computation into data-dependent (known as offline) and data-independent (known as online) phases. The former is dedicated to generating correlated randomness (e.g., Beaver triples), which are used later in the online phase reducing its computation time. Table~\ref{Computation-Results} shows a detailed overview of our protocol’s computational overhead, including CPU time and number of communication rounds. The evaluation is provided for both honest majority and dishonest majority with active or passive security. The protocol was evaluated using an arithmetic circuit for all security settings except for the dishonest majority and passive model, which according to our tests, is more efficient to be performed using mixed computations. Online-only benchmarks are also provided.

\begin{table}[t]
  \centering
  \footnotesize
  \caption{Computation Results (Time in Seconds)}
\label{Computation-Results}
\begin{tabular}{ p{0.05 \columnwidth} p{0.1 \columnwidth} p{0.09 \columnwidth} m{0.05 \columnwidth} m{0.07 \columnwidth} m{0.05  \columnwidth} m{0.07  \columnwidth} m{0.05  \columnwidth} m{0.07  \columnwidth} } 
\toprule
   \textbf{Users} & \multicolumn{2}{c}{\textbf{Security Model}}  &\multicolumn{2}{c}{\textbf{\shortstack{Base \\Protocol}}}  & \multicolumn{2}{c}{\textbf{\shortstack{Online\\ phase}}} &  \multicolumn{2}{c}{\textbf{\shortstack{Revealing\\ Deviations}}}\\
    &&& Time & Rounds & Time & Rounds & Time & Rounds\\
       \midrule
       \multirow{4}{*}{\textbf{1000}} & \multirow{2}{*}{\shortstack{Honest \\majority}} & Passive & 1.39 & 9129 & 1.20  & 9085 & 0.29 & 2085\\
       & & Active& 2.09 & 10229 & 1.40 & 10087 & 0.50 & 3092\\
        \noalign{\vspace{0.5ex}}\cline{2-9} \noalign{\vspace{0.5ex}}
      & \multirow{2}{*}{\shortstack{Dishonest\\majority}}& Passive & 9.70 & 44783  & 1.11 & 40004 & 0.29 & 8086 \\
      & & Active & 70.90 & 49838 & 1.80 & 40259 & 5.70 & 10410\\
      \midrule
      \multirow{4}{*}{\textbf{2000}} & \multirow{2}{*}{\shortstack{Honest \\majority}} & Passive & 2.70 & 18252 & 2.34 & 18168  & 0.56 & 4168\\
       & & Active & 3.80 & 20442 & 2.80 & 20170 & 0.97 & 6175\\
      \noalign{\vspace{0.5ex}}\cline{2-9} \noalign{\vspace{0.5ex}}
      & \multirow{2}{*}{\shortstack{Dishonest\\majority}}& Passive & 19.10 & 89481 & 2.30  & 80004 & 0.56 & 16108 \\
      & & Active & 142.70 & 99553 & 3.90 & 80508 & 9.00 & 20697\\
      \midrule
      \multirow{4}{*}{\textbf{3000}} & \multirow{2}{*}{\shortstack{Honest \\majority}} & Passive & 3.90 & 27380 & 3.50  & 27252 & 0.85 & 6252\\
       & & Active & 5.70 & 30663 & 4.20 & 30254 & 1.48 & 9259\\
        \noalign{\vspace{0.5ex}}\cline{2-9} \noalign{\vspace{0.5ex}}
      & \multirow{2}{*}{\shortstack{Dishonest\\majority}}& Passive & 28.80 &  134179 & 2.98 & 120004 & 0.81 & 24130 \\
      & & Active & 210.00 & 149271 & 5.60 & 120760 & 14.00 & 30987\\
      \midrule
      \multirow{4}{*}{\textbf{4000}} & \multirow{2}{*}{\shortstack{Honest \\majority}} & Passive & 5.20 & 36503  & 4.70 & 36335 & 1.09 & 8355\\
       & & Active & 7.50 & 40867  & 5.60 & 40337 & 1.89 & 12342\\
        \noalign{\vspace{0.5ex}}\cline{2-9} \noalign{\vspace{0.5ex}}
      & \multirow{2}{*}{\shortstack{Dishonest\\majority}}& Passive & 38.40 & 178877 & 4.01 & 160004 & 1.11 & 32152 \\
      & & Active & 279.50 &  198986 & 7.01 & 161009 & 17.96 & 41274\\
      \midrule
      \multirow{4}{*}{\textbf{5000}} & \multirow{2}{*}{\shortstack{Honest \\majority}} & Passive & 6.50 & 45630 & 5.80  & 45418 & 1.35 & 10418 \\
       & & Active & 9.40 & 51069 & 7.00 & 50420 & 2.40 & 15425\\
       \noalign{\vspace{0.5ex}}\cline{2-9} \noalign{\vspace{0.5ex}}
      & \multirow{2}{*}{\shortstack{Dishonest\\majority}}& Passive & 48.04  & 223575 & 5.01 & 200004 & 1.30 & 40174 \\
      & & Active & 351.00 & 248703 & 8.40 & 201258 & 22.40 & 51561 \\

\bottomrule

\end{tabular}
\end{table}

Our protocol is capable of handling 5000 users in less than ten seconds in the honest-majority setting, active model included. The dishonest majority, on the other hand, requires considerably more time because of the public key primitives it is based on. It takes around 50 seconds in the passive case and slightly less than 6 minutes in the malicious case because of the additional required steps such as MAC generation, oblivious transfer correlation checks and sacrificing. However, when the online phase is only considered, the results are clearly feasible to be applied in LEM billing even in the dishonest-majority setting, which is less than 9 seconds in all cases. In other words, after a trading period, users could receive their bills in a short time (suppliers do not need instant billing as they receive an aggregate of the bills). 

\begin{figure}[t]
\centering
  \includegraphics[width=\columnwidth]{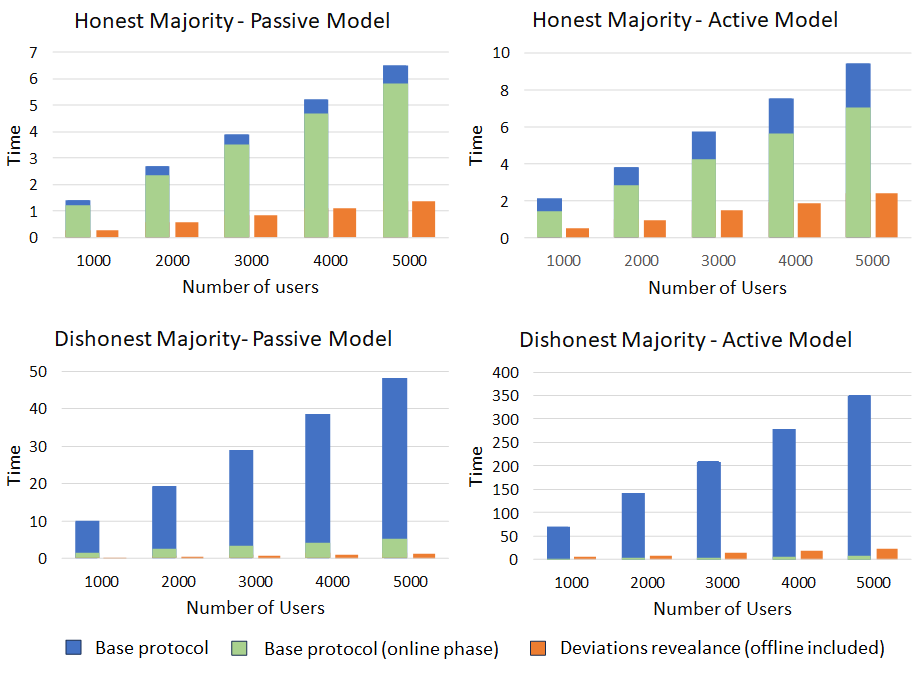}
   \captionsetup{justification=centering}
 \caption{Computational Results (Time in Seconds) }
 \label{PrivacyTechniques2}
 \end{figure}

Furthermore, the major overhead of our protocol is caused by the number of comparison operations executed for every user to check their individual deviations (Alg.~\ref{alg:IndividualBilling}). For example, in the honest-majority and passive security, eight interaction rounds are required per user. Due to this observation, we tested revealing individual users' deviations so that the individual comparisons could be conducted in clear. This would reveal some information about users, particularly whether they need to pay for the deviation cost, which is part of their bills. However, critical private data such as meter readings and bids' volumes cannot be inferred. The computation results of revealing individual deviations are shown in Table~\ref{Computation-Results}. A significant improvement can be easily noticed, in which the protocol takes less than 23 seconds for 5000 users in all different security settings, with the offline phase included.  
Figure~\ref{PrivacyTechniques2} visualises our results.

\section{Conclusions} \label{sec:conclusion}
In this work, we introduced a zone-based billing protocol for LEM based on MPC. The protocol considers imperfect bid fulfilment by splitting deviations cost amongst users while protecting their individual private data. We have analysed the complexity of our protocol in both honest-majority and dishonest-majority settings. The results show the feasibility of our billing protocol, as it can be performed for 5000 users in less than 9 seconds in the online phase for both security settings.


%





\ifCLASSOPTIONcaptionsoff
  \newpage
\fi

\bibliographystyle{IEEEtran}
\footnotesize
\bibliography{Zone_Based_Privacy_Preserving_Billing_for_Local_Energy_Market_Based_on_Multiparty_Computation}

\end{document}